\documentstyle[preprint,eqsecnum,aps]{revtex}
\begin{document}
\draft
\title{Quantum Aharonov-Bohm Billiard System}
\author{Der-San Chuu and De-Hone Lin \thanks{%
e-mail:d793314@phys.nthu.edu.tw}}
\address{Institute of Electro-Physics, National Chiao Tung University \\
Hsinchu 30043, Taiwan}
\date{\today}
\maketitle
\begin{abstract}
The Green's functions of the two and three-dimensional relativistic
Aharonov-Bohm (A-B) systems are given by the path integral approach. In
addition the exact radial Green's functions of the spherical A-B quantum
billiard system in two and three-dimensional are obtained via the
perturbation techanique of $\delta $-function.
\end{abstract}
\newpage \tolerance=10000
\section{Introduction}
It was points out in 1959 by A-B \cite{1} that the motion of a charged
particle can be affected by magnetic fields in regions from which the
particle is excluded. This interesting property of magnetic fields and thus
the vector potential ${\bf A(x)}$ in quantum mechanics has been well
confirmed experimentally. Because it raises some fundamental questions, the
A-B effect has much debated and written about over the years \cite{1e}.
Recently, the A-B billiard systems have been much interest in the contexts
of \ mesoscope \cite{1f,1g,1h,1c,1d}, nonlinear, and semi-classical dynamics 
\cite{1a,1b}.

In this contribution, we present the Green's functions of the relativistic 2
and 3-dimensional A-B system by the path integral approach. Furthermore, the
exact Green's functions of 2 and 3-dimensional relativistic spherical
quantum billiard with a singular magnetic flux (A-B magnetic field) centered
at the origin is given by the closed formula of the perturbation technique.
It is found that the energy spectra is only determined by the modified
Bessel function involving the partial wave Green's function of the
unperturbed relativistic A-B effect.

\section{Green's Function of the two-dimensional A-B Billiard System}

The starting point is the path integral representation for the Green's
function of a relativistic particle in external electromagnetic fields \cite
{2,3}: 
\begin{equation}
G({\bf x}_{b},{\bf x}_{a};E)=\frac{i\hbar }{2mc}\int_{0}^{\infty }dL\int 
{\cal D}\rho (\lambda )\Phi \left[ \rho (\lambda )\right] \int {\cal D}%
^{D}x(\lambda )\exp \left\{ -A_{E}\left[ {\bf x},{\bf \dot{x}}\right] /\hbar
\right\} \rho (0)  \label{a1}
\end{equation}
with the action 
\begin{equation}
A_{E}\left[ {\bf x},{\bf \dot{x}}\right] =\int_{\lambda _{a}}^{\lambda
_{b}}d\lambda \left[ \frac{m}{2\rho \left( \lambda \right) }{\bf \dot{x}}%
^{2}\left( \lambda \right) -i(e/c){\bf A(x)\cdot \dot{x}(}\lambda {\bf )}%
-\rho (\lambda )\frac{\left( E-V({\bf x})\right) ^{2}}{2mc^{2}}+\rho \left(
\lambda \right) \frac{mc^{2}}{2}\right] ,  \label{a2}
\end{equation}
where $L$ is defined as 
\begin{equation}
L=\int_{\lambda _{a}}^{\lambda _{b}}d\lambda \rho (\lambda ),  \label{a3}
\end{equation}
in which $\rho (\lambda )$ is an arbitrary dimensionless fluctuating scale
variable, $\rho (0)$ is the terminal point of the function $\rho (\lambda )$%
, and $\Phi \lbrack \rho (\lambda )]$ is some convenient gauge-fixing
functional \cite{2,3,4}. The only condition on $\Phi \lbrack \rho (\lambda
)] $ is that 
\begin{equation}
\int {\cal D}\rho (\lambda )\Phi \left[ \rho (\lambda )\right] =1.
\label{a4}
\end{equation}
$\hbar /mc$ is the well-known Compton wave length of a particle of mass $m$, 
${\bf A(x)}$ and $V({\bf x})$ stand for the vector and scalar potential of
the systems, respectively. $E$ is the system energy, and ${\bf {x}}$ is the
spatial part of the ($D+1$) vector $x^{\mu }=({\bf {x}},\tau )$.

For the pure Aharonov-Bohm (AB) system under consideration, the scalar
potential $V({\bf x})=0$ and the vector potential reads 
\begin{equation}
{\bf A(x)=}2g\frac{-y\hat{e}_{1}+x\hat{e}_{2}}{x^{2}+y^{2}},  \label{a5}
\end{equation}
where $\hat{e}_{1,2}$ stand for the unit vector along the $x,y$ axis,
respectively. In this problem, the functional $\Phi \left[ \rho (\lambda )%
\right] $ can be taken as the $\delta $-functional $\delta \left[ \rho -1%
\right] $ to fixed the value of $\rho (\lambda )$ to unity \cite{3}. For
convenience, Let's now introduce the azimuthal angle around the magnetic
tube: 
\begin{equation}
\varphi ({\bf x})=\arctan (y/x).  \label{a6}
\end{equation}
The components of the vector potential can be therefore expressed as 
\begin{equation}
A_{i}=2g\partial _{i}\varphi ({\bf x}).  \label{a7}
\end{equation}
The associated magnetic field lines are confined to an infinitely thin tube
along the z-axis: 
\begin{equation}
B_{3}=2g\epsilon _{3ij}\partial _{i}\partial _{j}\varphi ({\bf x})=4\pi
g\delta ({\bf x}_{\bot }),  \label{a8}
\end{equation}
where ${\bf x}_{\bot }$ stands for the transverse vector ${\bf x}_{\bot
}=(x,y)$. Note that the derivatives in front of $\varphi ({\bf x})$ commute
everywhere, except at the origin where Stokes' theorem yields 
\begin{equation}
\int d^{2}x\left( \partial _{1}\partial _{2}-\partial _{2}\partial
_{1}\right) \varphi ({\bf x})=\oint d\varphi =2\pi .  \label{a9}
\end{equation}
The magnetic flux through the tube is defined by the integral 
\begin{equation}
\Omega =\int d^{2}xB_{3}.  \label{a10}
\end{equation}
This shows that the coupling constant $g$ is related to the magnetic flux by 
\begin{equation}
g=\frac{\Omega }{4\pi }.  \label{a11}
\end{equation}
Inserting $A_{i}=2g\partial _{i}\varphi ({\bf x})$ in Eq. (\ref{a2}), the
magnetic interaction takes the form 
\begin{equation}
A_{{\rm mag}}=-\hbar \beta _{0}\int_{0}^{L}d\lambda \dot{\varphi}(\lambda ),
\label{a12}
\end{equation}
where $\varphi (\lambda )=\varphi ({\bf x}(\lambda )),$ and $\beta _{0}$ is
the dimensionless number 
\begin{equation}
\beta _{0}=-\frac{2eg}{\hbar c}.  \label{a13}
\end{equation}
The minus sign is a matter of convention. Since the particle orbits are
present at all times, their worldlines in spacetime can be considered as
being closed at infinity, and the integral 
\begin{equation}
n=\frac{1}{2\pi }\int_{0}^{L}d\lambda \dot{\varphi}(\lambda )  \label{a14}
\end{equation}
is the topological invariant with integer values of the winding number $n$.
The magnetic interaction is therefore purely topological, its value being%
\label{1} 
\begin{equation}
A_{{\rm mag}}=-\hbar \beta _{0}2n\pi .  \label{a15}
\end{equation}
After adding this to the action of Eq. (\ref{a2}) in the radial
decomposition of the relativistic path integral, we rewrite the sum over the
azimuthal quantum number via Poisson's summation formula 
\begin{equation}
\sum_{k=-\infty }^{\infty }f(k)=\int_{-\infty }^{\infty }dy\sum_{n=-\infty
}^{\infty }e^{2\pi nyi}f(y).  \label{a16}
\end{equation}
This leads to 
\begin{equation}
G({\bf x}_{b},{\bf x}_{a};E)=\frac{i\hbar }{2mc}\int_{0}^{\infty }dLe^{L%
{\cal E}/\hbar }\int_{-\infty }^{\infty }d\alpha K\left(
r_{b},r_{a};L\right) _{\alpha }\cdot \sum_{n=-\infty }^{\infty }\frac{1}{%
2\pi }e^{i\left( \alpha -\beta _{0}\right) \left( \varphi _{b}+2n\pi
-\varphi _{a}\right) },  \label{a17}
\end{equation}
where the pseudoenergy ${\cal E=}\left( E^{2}-m^{2}c^{4}\right) /2mc^{2},$
and the radial pseudopropagator $K\left( r_{b},r_{a};L\right) _{\alpha }$
has the representation 
\begin{equation}
K\left( r_{b},r_{a};L\right) _{\alpha }=\frac{m}{\hbar }\frac{1}{L}%
e^{-m\left( r_{b}^{2}+r_{a}^{2}\right) /2\hbar L}I_{\alpha }\left( \frac{%
mr_{b}r_{a}}{\hbar L}\right) .  \label{a18}
\end{equation}
The sum over all $n$ in Eq. (\ref{a17}) forces $\alpha $ to be equal to $%
\beta _{0}$ modulo an arbitrary integral number. The result is 
\begin{equation}
G({\bf x}_{b},{\bf x}_{a};E)=\frac{i\hbar }{2mc}\int_{0}^{\infty }dLe^{L%
{\cal E}/\hbar }K\left( {\bf x}_{b},{\bf x}_{a};L\right)  \label{a19}
\end{equation}
in which $K\left( {\bf x}_{b},{\bf x}_{a};L\right) $ is given by 
\begin{equation}
K\left( {\bf x}_{b},{\bf x}_{a};L\right) =\sum_{n=-\infty }^{\infty }K\left(
r_{b},r_{a};L\right) _{n+\beta _{0}}\frac{1}{2\pi }e^{in\left( \varphi
_{b}-\varphi _{a}\right) }.  \label{a20}
\end{equation}
From Eq. (\ref{a19}), we observe that $K\left( {\bf x}_{b},{\bf x}%
_{a};L\right) $ can be viewed as the propagator of AB system with the
pseudoenergy ${\cal E}$. The entire Green's function can be obtained by
doing the integration. At this place, let us first discuss the wave function
properties by noting the relation 
\begin{equation}
\Psi \left( r,\varphi ;L\right) =\int_{0}^{\infty }r_{0}dr_{0}\int_{-\pi
}^{\pi }d\varphi _{0}K\left( {\bf x},{\bf x}_{0};L\right) \Psi \left(
r_{0},\varphi _{0};0\right) ,  \label{a21}
\end{equation}
where $\Psi \left( r_{0},\varphi _{0};0\right) $ reads 
\begin{equation}
\Psi \left( r_{0},\varphi _{0};0\right) =e^{-ikr_{0}\cos \varphi
_{0}}e^{-i\beta _{0}\varphi _{0}}  \label{a22}
\end{equation}
with the $k=\sqrt{E^{2}-m^{2}c^{4}}/\hbar c$. Inserting the expression of
Eq. (\ref{a20}) in Eq. (\ref{a21}), we obtain 
\[
\Psi \left( r,\varphi ;L\right) =\sum_{n=-\infty }^{\infty }e^{in\varphi
}\int_{0}^{\infty }r_{0}dr_{0}\frac{m}{\hbar }\frac{1}{L}e^{-m\left(
r^{2}+r_{0}^{2}\right) /2\hbar L}I_{n+\beta _{0}}\left( \frac{mrr_{0}}{\hbar
L}\right) 
\]
\begin{equation}
\times \frac{1}{2\pi }\int_{-\pi }^{\pi }d\varphi _{0}e^{-ikr_{0}\cos
\varphi _{0}-i\left( n+\beta _{0}\right) \varphi _{0}}.  \label{a23}
\end{equation}
The angular integration can be performed by using the formula 
\begin{equation}
\lim_{z\rightarrow \infty }\frac{1}{2\pi }\int_{-\pi }^{\pi }d\theta
e^{-iz\cos \theta -i\nu \theta }=I_{\nu }(-iz).  \label{a24}
\end{equation}
We arrive at 
\begin{equation}
\Psi \left( r,\varphi ;L\right) =\sum_{n=-\infty }^{\infty }e^{in\varphi
}\int_{0}^{\infty }r_{0}dr_{0}\frac{m}{\hbar }\frac{1}{L}e^{-m\left(
r^{2}+r_{0}^{2}\right) /2\hbar L}I_{n+\beta _{0}}\left( \frac{mrr_{0}}{\hbar
L}\right) I_{n+\beta _{0}}\left( -ikr_{0}\right) .  \label{a25}
\end{equation}
The integral can perform with the help of the formula 
\begin{equation}
\int_{0}^{\infty }drre^{-r^{2}/a}I_{\nu }(\varsigma r)I_{\nu }(\xi r)=\frac{a%
}{2}e^{a(\xi ^{2}+\varsigma ^{2})/4}I_{\nu }\left( \frac{a\xi \varsigma }{2}%
\right) .  \label{a26}
\end{equation}
It yields 
\begin{equation}
\Psi \left( r,\varphi ;L\right) =\sum_{n=-\infty }^{\infty }e^{in\varphi
}I_{n+\beta _{0}}\left( -ikr\right) e^{-(\hbar k)^{2}L/2m\hbar }.
\label{a27}
\end{equation}
Extracting the wave function depending only on spatial variable $\left(
r,\varphi \right) $, we obtain 
\begin{equation}
\Psi \left( r,\varphi \right) =\sum_{n=-\infty }^{\infty }\left( -i\right)
^{n+\beta _{0}}J_{n+\beta _{0}}\left( kr\right) e^{in\varphi },  \label{a28}
\end{equation}
where the equality $I_{\nu }(-iz)=(-i)^{\nu }J_{\nu }(z)$ has been used.
This wave function coincides with the local approach in the relativistic
version. The scattering wave of the relativistic AB effect can be extracted
by spliting the wave function into three parts: 
\[
\Psi =\Psi _{1}+\Psi _{2}+\Psi _{3} 
\]
\begin{equation}
=\sum_{n=1}^{\infty }\left( -i\right) ^{n+\beta _{0}}J_{n+\beta _{0}}\left(
kr\right) e^{in\varphi }+\sum_{n=1}^{\infty }\left( -i\right) ^{n-\beta
_{0}}J_{n-\beta _{0}}\left( kr\right) e^{-in\varphi }+\left( -i\right)
^{\beta _{0}}J_{\beta _{0}}\left( kr\right) e^{in\varphi }.  \label{a29}
\end{equation}
It is not difficult to calculate the asymptotic behavior is given by 
\begin{equation}
\Psi =\Psi _{1}+\Psi _{2}+\Psi _{3}\stackrel{r\rightarrow \infty }{%
\longrightarrow }e^{-ikr\cos \varphi }e^{-i\beta _{0}\varphi }+f(\varphi )%
\frac{e^{ikr}}{\sqrt{kr}}  \label{a30}
\end{equation}
with the scattering amplitude 
\begin{equation}
f(\varphi )=\frac{e^{-i3\pi /4}}{\sqrt{2\pi }}\sin \pi \beta _{0}\frac{%
e^{-i\varphi /2}}{\cos \left( \varphi /2\right) }.  \label{a31}
\end{equation}
The corresponding cross section is 
\begin{equation}
\frac{d\sigma }{d\varphi }=\frac{\sin ^{2}(\pi \beta _{0})}{2\pi }\frac{1}{%
\cos ^{2}(\varphi /2)}.  \label{a32}
\end{equation}
It has a strong peak near the forward direction $\varphi \approx \pi $. For $%
\beta _{0}=$ integer, there is no scattering at all. This concludes the
discussion of the wave function by the path integral solution of the
relativistic spinless AB system.

Now let's begin to discuss the A-B billiard system. With a method developed
in Ref. \cite{5,5a}, the exact Green's function of quantum billiard for a
spherically shaped, impenetrable wall located at the radius $r=a$ is given
by the following formula 
\begin{equation}
G^{({\rm wall})}(r_{b},r_{a};E)=G(r_{b},r_{a};E)-\frac{%
G(r_{b},a;E)G(a,r_{a};E)}{G(a,a;E)},  \label{a33}
\end{equation}
where $G(r_{b},r_{a};E)$ is the Green's function of unperturbed radial
propagator. For the pure A-B system under consideration, It can be obtained
by noting Eqs. (\ref{a19}) and (\ref{a20}) and reads 
\begin{equation}
G(r_{b},r_{a};E)=\int_{0}^{\infty }dLe^{L{\cal E}/\hbar }\frac{m}{\hbar }%
\frac{1}{L}e^{-m\left( r_{b}^{2}+r_{a}^{2}\right) /2\hbar L}I_{\left|
n+\beta _{0}\right| }\left( \frac{mr_{b}r_{a}}{\hbar L}\right) .  \label{a34}
\end{equation}
The integral can perform by using the integral representation [\cite{7},
p.200], 
\begin{equation}
\int_{0}^{\infty }\frac{dz}{z}e^{-pz-(a+b)/2z}I_{\nu }(\frac{a-b}{2z}%
)=2I_{\nu }\left( \sqrt{p}(\sqrt{a}-\sqrt{b})\right) K_{\nu }\left( \sqrt{p}(%
\sqrt{a}+\sqrt{b})\right) ,  \label{a35}
\end{equation}
yielding 
\[
G(r_{b},r_{a};E)=\frac{2m}{\hbar }I_{\left| n+\beta _{0}\right| }\left( 
\sqrt{-m{\cal E}/2\hbar ^{2}}\left( r_{b}+r_{a}-\left| r_{b}-r_{a}\right|
\right) \right) 
\]
\begin{equation}
\times K_{\left| n+\beta _{0}\right| }\left( \sqrt{-m{\cal E}/2\hbar ^{2}}%
\left( r_{b}+r_{a}+\left| r_{b}-r_{a}\right| \right) \right) .  \label{a36}
\end{equation}
This gives the Green's function with a wall located at $r=a$ e.g. for $%
r_{a}<r_{b}<a:$%
\[
G^{({\rm wall})}(r_{b},r_{a};E)=\frac{2m}{\hbar }\left[ I_{\left| n+\beta
_{0}\right| }\left( \sqrt{-2m{\cal E}/\hbar ^{2}}a\right) K_{\left| n+\beta
_{0}\right| }\left( \sqrt{-2m{\cal E}/\hbar ^{2}}r_{b}\right) -\left(
K\leftrightarrow I\right) \right] 
\]
\begin{equation}
\times \frac{I_{\left| n+\beta _{0}\right| }\left( \sqrt{-2m{\cal E}/\hbar
^{2}}r_{a}\right) }{I_{\left| n+\beta _{0}\right| }\left( \sqrt{-2m{\cal E}%
/\hbar ^{2}}a\right) }.  \label{a37}
\end{equation}
The corresponding bound state energy spectra are given by the equation 
\begin{equation}
I_{\left| n+\beta _{0}\right| }\left( \sqrt{-2m{\cal E}/\hbar ^{2}}a\right)
=0.  \label{a38}
\end{equation}
We see that the presence of the flux line in the circular billiard simply
changes the order of the Bessel functions from the integer to fractional. If
we assume that $-\beta _{0}$ can take a continuous range of values between $%
0 $ and $1$, the symmetry $\left| \beta _{0}\right| \leftrightarrow
(1+\left| \beta _{0}\right| )$ in the quantum spectrum allows the
restriction to $0\leq \left| \beta _{0}\right| \leq 0.5$. For integer flux $%
\left| \beta _{0}\right| =0,1,2,\cdots ,$ the quantum spectrum is unaltered
by the flux line. This is seen from the fact that for any integer value of $%
\beta _{0}$ the angular momentum gets redefined and the new set is
isomorphic to the old one both in terms of the spectrum and eigenstates.
This mapping, however, has no classical analog since the classically allowed
angular momenta remain the same.

\section{Green's Function of the Three-dimensional A-B Billiard System}

For solving the path integral of A-B system in three-dimensional space,
Let's now introduce the space-time transformation \cite{8,9} 
\begin{equation}
\epsilon _{n}^{\lambda }=\epsilon _{n}^{s}f({\bf x}_{n})  \label{b1}
\end{equation}
with the short ``time'' interval $\epsilon _{n}^{\lambda }=\lambda
_{n}-\lambda _{n-1}$ to regularize the path integral in time-sliced form for
getting a tractable one. Then we have the regularization path integral as
following: 
\begin{equation}
G({\bf x}_{b},{\bf x}_{a};E)\approx \frac{i\hbar }{2mc}\int_{0}^{\infty }dS%
\frac{f({\bf {x}}_{a})}{\left( \frac{2\pi \hbar \epsilon _{b}^{s}f({\bf {x}}%
_{b})}{m}\right) ^{3/2}}\prod_{n=1}^{N}\left[ \int_{-\infty }^{\infty }\frac{%
d^{3}x_{n}}{\left( \frac{2\pi \hbar \epsilon _{n}^{s}f({\bf {x}}_{n})}{m}%
\right) ^{3/2}}\right] \exp \left\{ -\frac{1}{\hbar }A^{N}\right\}
\label{b2}
\end{equation}
with the $s$-sliced action 
\begin{equation}
A^{N}=\sum_{n=1}^{N+1}\left[ \frac{m\left( {\bf {x}}_{n}-{\bf {x}}%
_{n-1}\right) ^{2}}{2\epsilon _{n}^{s}f({\bf {x}}_{n})}-i\frac{e}{c}{\bf A}(%
{\bf x}_{n})\cdot ({\bf x}_{n}-{\bf x}_{n-1})~-\epsilon _{n}^{s}f({\bf {x}}%
_{n})\frac{E^{2}}{2mc^{2}}+\epsilon _{n}^{s}f({\bf {x}}_{n})\frac{mc^{2}}{2}%
\right] .  \label{b3}
\end{equation}
The sign $\approx $ in Eq. (\ref{b2}) becomes an equality for $N\rightarrow
\infty $. The regularization $f({\bf {x)}}$ in this problem can be chosen as
the radial distance $r=\sqrt{x^{2}+y^{2}+z^{2}}$ and thus $f({\bf {x}}%
_{a})=r_{a}$. Since we shall introduce the KS-transformation \cite{8}, let
us insert in Eq. (\ref{b2}) a functional integral representation of unity 
\begin{equation}
\prod_{n=1}^{N+1}\left[ \int \frac{d\triangle w_{n}}{\left( 2\pi \hbar
\epsilon _{n}^{s}\rho _{n}r_{n}/m\right) ^{1/2}}\right] \exp \left\{ -\frac{1%
}{\hbar }\sum_{n=1}^{N+1}\frac{m}{2}\frac{(\triangle w_{n})^{2}}{\epsilon
_{n}^{s}\rho _{n}r_{n}}\right\} =1.  \label{b4}
\end{equation}
The $w_{n}$ is a fictitious fourth coordinate axis. With this, the path
integral of 3-dimensional A-B effect can be rewritten as the 4-dimensional
path integral in the time-sliced version 
\begin{equation}
G({\bf x}_{b},{\bf x}_{a};E)\approx \frac{i\hbar }{2mc}\int_{0}^{\infty
}dS\int dw_{a}\frac{r_{a}}{\left( \frac{2\pi \hbar \epsilon _{b}^{s}r_{b}}{m}%
\right) ^{2}}\prod_{n=2}^{N+1}\left[ \int_{-\infty }^{\infty }\frac{%
d\triangle \vec{x}_{n}}{\left( \frac{2\pi \hbar \epsilon _{n}^{s}r_{n-1}}{m}%
\right) ^{2}}\right] \exp \left\{ -\frac{1}{\hbar }A^{N}\right\}  \label{b5}
\end{equation}
with the sliced action 
\begin{equation}
A^{N}=\sum_{n=1}^{N+1}\left[ \frac{m\left( \vec{x}_{n}-\vec{x}_{n-1}\right)
^{2}}{2\epsilon _{n}^{s}r_{n}}-i(e/c){\bf A}({\bf x}_{n})\cdot ({\bf x}_{n}-%
{\bf x}_{n-1})-\epsilon _{n}^{s}r_{n}\frac{E^{2}}{2mc^{2}}+\epsilon
_{n}^{s}r_{n}\frac{mc^{2}}{2}\right] ,  \label{b6}
\end{equation}
where the kinetic term in Eq. (\ref{b3}) has been replaced with the
four-vector $\vec{x}$ due to the reason of Eq. (\ref{b4}) and the integrals
over $\triangle \vec{x}_{n}$ may be performed successively from $n=N$ down
to $n=1$. The notation's change of the measure of integration is necessary
for discussing the path integral in curved space, since $\vec{x}_{n}$ are
Cartesian coordinates and certainly identical in the time-sliced expressions 
\cite{9}: 
\begin{equation}
\prod_{n=1}^{N}\left[ \int_{-\infty }^{\infty }d\vec{x}_{n}\right]
=\prod_{n=2}^{N+1}\left[ \int_{-\infty }^{\infty }d\triangle \vec{x}_{n}%
\right] .  \label{b7}
\end{equation}
However, their images under a non-holonomic mapping are different so that
the initial form of the sliced path integral is a matter of choice. In the
space with curvature and torsion it has been proved in Ref. \cite{9}. Only
the right-hand side of Eq. (\ref{b7}) gives the properly correct results. To
go further, let's adjust for having the same time-sliced index the measure
via the following approximation 
\begin{equation}
\frac{r_{a}}{\left( \frac{2\pi \hbar \epsilon _{b}^{s}r_{b}}{m}\right) ^{2}}%
\prod_{n=2}^{N+1}\left[ \int_{-\infty }^{\infty }\frac{d\triangle \vec{x}_{n}%
}{\left( \frac{2\pi \hbar \epsilon _{n}^{s}r_{n-1}}{m}\right) ^{2}}\right]
\approx \frac{1}{r_{a}}\frac{1}{\left( \frac{2\pi \hbar \epsilon _{b}^{s}}{m}%
\right) ^{2}}\prod_{n=2}^{N+1}\left[ \int_{-\infty }^{\infty }\frac{%
d\triangle \vec{x}_{n}}{\left( \frac{2\pi \hbar \epsilon _{n}^{s}r_{n}}{m}%
\right) ^{2}}\right] .  \label{b8}
\end{equation}
With the help of the approximation, we arrive at 
\begin{equation}
G({\bf x}_{b},{\bf x}_{a};E)\approx \frac{i\hbar }{2mc}\int_{0}^{\infty
}dS\int \frac{dw_{a}}{r_{a}}\frac{1}{\left( \frac{2\pi \hbar \epsilon
_{b}^{s}}{m}\right) ^{2}}\prod_{n=2}^{N+1}\left[ \int_{-\infty }^{\infty }%
\frac{d\triangle \vec{x}_{n}}{\left( \frac{2\pi \hbar \epsilon _{n}^{s}r_{n}%
}{m}\right) ^{2}}\right] \exp \left\{ -\frac{1}{\hbar }A^{N}\right\} .
\label{b9}
\end{equation}
This path integral can be simplified by the KS-transformation 
\begin{equation}
d\vec{x}=2A(\vec{u})d\vec{u}.  \label{b10}
\end{equation}
The $4\times 4$ matrix $A(\vec{u})$ is chosen as 
\begin{equation}
A(\vec{u})=\left( 
\begin{array}{cccc}
u^{3} & u^{4} & u^{1} & u^{2} \\ 
u^{4} & -u^{3} & -u^{2} & u^{1} \\ 
u^{1} & u^{2} & -u^{3} & -u^{4} \\ 
u^{2} & -u^{1} & u^{4} & -u^{3}
\end{array}
\right) .  \label{b11}
\end{equation}
The transformations of the volume element and velocity are given as 
\begin{equation}
d\vec{x}=16r^{2}d\vec{u},  \label{b12}
\end{equation}
\begin{equation}
\vec{x}^{\prime 2}=4\vec{u}^{2}\vec{u}^{\prime 2}=4r\vec{u}^{\prime 2}.
\label{b13}
\end{equation}
Furthermore, the magnetic interaction under the KS-transformation turns into 
\begin{equation}
{\bf A(x)\cdot \dot{x}(}\lambda {\bf )=}-2g\frac{y\dot{x}-x\dot{y}}{%
x^{2}+y^{2}}=-2g\left[ \frac{u^{1}\dot{u}^{2}-u^{2}\dot{u}^{1}}{\left(
u^{1}\right) ^{2}+\left( u^{2}\right) ^{2}}+\frac{u^{4}\dot{u}^{3}-u^{3}\dot{%
u}^{4}}{\left( u^{3}\right) ^{2}+\left( u^{4}\right) ^{2}}\right]
\label{b14}
\end{equation}
or in the time-sliced version 
\[
{\bf A}({\bf x}_{n})\cdot ({\bf x}_{n}-{\bf x}_{n-1})=-2g\frac{%
y_{n}\triangle x_{n}-x_{n}\triangle y_{n}}{x_{n}^{2}+y_{n}^{2}} 
\]
\begin{equation}
=-2g\left[ \frac{u_{n}^{1}\triangle u_{n}^{2}-u_{n}^{2}\triangle u_{n}^{1}}{%
\left( u_{n}^{1}\right) ^{2}+\left( u_{n}^{2}\right) ^{2}}+\frac{%
u_{n}^{4}\triangle u_{n}^{3}-u_{n}^{3}\triangle u_{n}^{4}}{\left(
u_{n}^{3}\right) ^{2}+\left( u_{n}^{4}\right) ^{2}}\right] .  \label{b15}
\end{equation}
We obtain a path integral equivalent to Eq. (\ref{b9}) 
\begin{equation}
G({\bf x}_{b},{\bf x}_{a};E)=\frac{i\hbar }{2mc}\int_{0}^{\infty }dS\ G(\vec{%
u}_{b},\vec{u}_{a};S),  \label{b16}
\end{equation}
where $G(\vec{u}_{b},\vec{u}_{a};S)$ denotes the s-sliced amplitude 
\begin{equation}
G(\vec{u}_{b},\vec{u}_{a};S)\approx \frac{1}{16}\int \frac{dw_{a}}{r_{a}}%
\frac{1}{\left( \frac{2\pi \hbar \epsilon _{b}^{s}}{M}\right) ^{2}}%
\prod_{n=1}^{N}\left[ \int_{-\infty }^{\infty }\frac{d\vec{u}_{n}}{\left( 
\frac{2\pi \hbar \epsilon _{n}^{s}}{M}\right) ^{2}}\right] \exp \left\{ -%
\frac{1}{\hbar }A^{N}\right\}  \label{b17}
\end{equation}
with the action 
\begin{equation}
A^{N}=\sum_{n=1}^{N+1}\left\{ \frac{M(\triangle \vec{u}_{n})^{2}}{2\epsilon
_{n}^{s}}-i(e/c)\left[ \vec{A}(u_{n})\cdot \triangle \vec{u}_{n}\right]
+\epsilon _{n}^{s}\frac{M\omega ^{2}\vec{u}_{n}^{2}}{2}\right\} .
\label{b18}
\end{equation}
Here 
\begin{equation}
M=4m,\quad \omega ^{2}=\frac{m^{2}c^{4}-E^{2}}{4m^{2}c^{2}},  \label{b19}
\end{equation}
and 
\begin{equation}
\vec{A}(u_{n})\cdot \triangle \vec{u}_{n}=-2g\left[ \frac{u_{n}^{1}\triangle
u_{n}^{2}-u_{n}^{2}\triangle u_{n}^{1}}{\left( u_{n}^{1}\right) ^{2}+\left(
u_{n}^{2}\right) ^{2}}+\frac{u_{n}^{4}\triangle u_{n}^{3}-u_{n}^{3}\triangle
u_{n}^{4}}{\left( u_{n}^{3}\right) ^{2}+\left( u_{n}^{4}\right) ^{2}}\right]
.  \label{b20}
\end{equation}
In the continuum limit, this amounts to 
\begin{equation}
G({\bf x}_{b},{\bf x}_{a};E)=\frac{i\hbar }{2mc}\int_{0}^{\infty }dS\ \frac{1%
}{16}\int \frac{dw_{a}}{r_{a}}\int {\cal D}\vec{u}(s)\exp \left\{ -\frac{1}{%
\hbar }A^{N}\right\}  \label{b21}
\end{equation}
with the action 
\begin{equation}
A=\int_{0}^{S}ds\left[ \frac{M\vec{u}^{\prime 2}}{2}-2i(e/c)(\vec{A}(s)\cdot 
\vec{u}^{\prime }(s))+\frac{M\omega ^{2}\vec{u}^{2}}{2}\right] .  \label{b22}
\end{equation}
There are no $s$-slicing corrections. This is ensured by the affine
connection of KS-transformation satisfying 
\begin{equation}
\Gamma _{\mu }^{\;\;\;\mu \lambda }=g^{\mu \nu }e_{i}^{\;\lambda }\partial
_{\mu }e_{\;\nu }^{i}=0  \label{b23}
\end{equation}
with the basis triads $e_{\;\nu }^{i}(q)=\partial x^{i}/\partial q^{\mu }$
and the transverse gauge $\partial _{\mu }A^{\mu }=0$ \cite{9}. Note that
the system becomes separable like $R^{4}\rightarrow R^{2}\times R^{2}$ in
which each $R^{2}$ has a dynamical model a 2-dimensional simple harmonic
oscillator moving in the A-B magnetic fields. Its Green's function is given
by \cite{8,9} 
\begin{equation}
\frac{M\omega }{\hbar \sinh \omega s}\sum_{k=-\infty }^{\infty
}e^{ik(\varphi _{b}-\varphi _{a})}\exp \left\{ -\frac{M\omega }{2\hbar }%
\left( \sigma _{b}^{2}+\sigma _{a}^{2}\right) \coth \omega s\right\} I_{\mid
k+\beta _{0}\mid }\left( \frac{M}{\hbar }\frac{\omega \sigma _{b}\sigma _{a}%
}{\sinh \omega s}\right) ,  \label{b24}
\end{equation}
where $\sigma =\sqrt{x^{2}+y^{2}}$ is the radial length$,I_{\nu }$ is the
modified Bessel function, and $\beta _{0}\equiv -2eg/\hbar c$. Therefore, we
obtain the entire Green's function 
\[
G({\bf x}_{b},{\bf x}_{a};E)=\frac{i\hbar }{2mc}\int_{0}^{\infty }dS\ \frac{1%
}{16}\int \frac{dw_{a}}{r_{a}}\left( \frac{M\omega }{\hbar \sinh \omega s}%
\right) ^{2} 
\]
\[
\times \sum_{k_{1}=-\infty }^{\infty }\sum_{k_{2}=-\infty }^{\infty
}e^{ik_{1}(\varphi _{1,b}-\varphi _{1,a})}e^{ik_{2}(\varphi _{2,b}-\varphi
_{2,a})} 
\]
\[
\times \exp \left\{ -\frac{M\omega }{2\hbar }\left( \sigma _{1,b}^{2}+\sigma
_{1,a}^{2}+\sigma _{2,b}^{2}+\sigma _{2,a}^{2}\right) \coth \omega s\right\} 
\]
\begin{equation}
\times I_{\mid k_{1}+\beta _{0}\mid }\left( \frac{M}{\hbar }\frac{\omega
\sigma _{1,b}\sigma _{1,a}}{\sinh \omega s}\right) I_{\mid k_{2}+\beta
_{0}\mid }\left( \frac{M}{\hbar }\frac{\omega \sigma _{2,b}\sigma _{2,a}}{%
\sinh \omega s}\right) .  \label{b25}
\end{equation}
Here the coordinate transformations $(\sigma _{1},\varphi _{1})$ and $%
(\sigma _{2},\varphi _{2})$ are defined as 
\begin{equation}
\left. 
\begin{array}{c}
u^{1}=\sigma _{1}\sin \varphi _{1} \\ 
u^{2}=\sigma _{1}\cos \varphi _{1} \\ 
u^{3}=\sigma _{2}\cos \varphi _{2} \\ 
u^{4}=\sigma _{2}\sin \varphi _{2}
\end{array}
\right\} .  \label{b26}
\end{equation}
To perform the $w_{a}$ integration, let's express the variables $(\sigma
_{1},\varphi _{1},\sigma _{2},\varphi _{2})$ in terms of the Euler angle
variables by defining: 
\begin{equation}
\left. 
\begin{array}{l}
u^{1}=\sqrt{r}\cos (\theta /2)\cos \left[ (\varphi +\gamma )/2\right] \\ 
u^{2}=-\sqrt{r}\cos (\theta /2)\sin \left[ (\varphi +\gamma )/2\right] \\ 
u^{3}=\sqrt{r}\sin (\theta /2)\cos \left[ (\varphi -\gamma )/2\right] \\ 
u^{4}=\sqrt{r}\sin (\theta /2)\sin \left[ (\varphi -\gamma )/2\right]
\end{array}
\right\} \qquad \left( 
\begin{array}{c}
0\leq \theta \leq \pi \\ 
0\leq \varphi \leq 2\pi \\ 
0\leq \gamma \leq 4\pi
\end{array}
\right)  \label{b27}
\end{equation}
and identify

\begin{equation}
\left. 
\begin{array}{l}
\sigma _{1}=\sqrt{r}\cos (\theta /2) \\ 
\varphi _{1}=(\varphi +\gamma +\pi )/2 \\ 
\sigma _{2}=\sqrt{r}\sin (\theta /2) \\ 
\varphi _{2}=(\varphi -\gamma )/2
\end{array}
\right\} .  \label{b28}
\end{equation}
Then one can change the $w_{a}$-integration into the $\gamma _{a}$%
-integration whose result is easily represented as the Kronecker delta $%
\delta _{k_{1},k_{2}}$. Hence, we carry out $k_{2}$-summation and finally
becomes 
\[
G({\bf x}_{b},{\bf x}_{a};E)=\frac{i\hbar }{2mc}\frac{m^{2}\omega }{\pi
\hbar ^{2}}\sum_{k=-\infty }^{\infty }e^{ik(\varphi _{b}-\varphi _{a})} 
\]
\[
\times \int_{0}^{\infty }d\eta \frac{1}{\sinh ^{2}\eta }e^{-\frac{M\omega }{%
2\hbar }(r_{b}+r_{a})\coth \eta } 
\]
\begin{equation}
\times I_{\mid k+\beta _{0}\mid }\left( \frac{M\omega \sqrt{r_{b}r_{a}}}{%
\hbar \sinh \eta }\cos \theta _{b}/2\cos \theta _{a}/2\right) I_{\mid
k+\beta _{0}\mid }\left( \frac{M\omega \sqrt{r_{b}r_{a}}}{\hbar \sinh \eta }%
\sin \theta _{b}/2\sin \theta _{a}/2\right) ,  \label{b29}
\end{equation}
where we have defined the new variable $\eta =\omega s$. The product of
modified Bessel functions can simplify by making use of the addition theorem 
\cite{3} 
\[
I_{\nu }\left( z\sin \alpha /2\sin \beta /2\right) I_{\mu }\left( z\cos
\alpha /2\cos \beta /2\right) 
\]
\[
=\frac{2}{z}\left( \sin \alpha /2\sin \beta /2\right) ^{\nu }\left( \cos
\alpha /2\cos \beta /2\right) ^{\mu }\sum_{l=0}^{\infty }\frac{l!\Gamma
\left( l+\mu +\nu +1\right) \left( 2l+\mu +\nu +1\right) }{\Gamma \left(
l+\mu +1\right) \Gamma \left( l+\nu +1\right) } 
\]
\begin{equation}
\times I_{2l+\mu +\nu +1}\left( z\right) P_{l}^{(\mu ,\nu )}\left( \cos
\theta _{b}\right) P_{l}^{(\mu ,\nu )}\left( \cos \theta _{a}\right) ,
\label{b30}
\end{equation}
where $P_{l}^{(\mu ,\nu )}$ is Jacobi polynomial (e.g. p.209 \cite{9a}). The
Green's function in Eq. (\ref{b29}) becomes 
\[
G({\bf x}_{b},{\bf x}_{a};E)=\frac{i\hbar }{2mc}\frac{m}{2\pi \hbar \sqrt{%
r_{b}r_{a}}}\sum_{k=-\infty }^{\infty }\sum_{l=0}^{\infty }e^{ik(\varphi
_{b}-\varphi _{a})} 
\]
\[
\times \left( \cos \theta _{b}/2\cos \theta _{a}/2\sin \theta _{b}/2\sin
\theta _{a}/2\right) ^{\mid k+\beta _{0}\mid }\frac{l!\Gamma \left( l+2\mid
k+\beta _{0}\mid +1\right) \left( 2l+2\mid k+\beta _{0}\mid +1\right) }{%
\Gamma ^{2}\left( l+\mid k+\beta _{0}\mid +1\right) } 
\]
\[
\times \left\{ \vbox to 24pt{}\int_{0}^{\infty }d\eta \frac{1}{\sinh \eta }%
e^{-\frac{M\omega }{2\hbar }(r_{b}+r_{a})\coth \eta }I_{2l+2\mid k+\beta
_{0}\mid +1}\left( \frac{M\omega \sqrt{r_{b}r_{a}}}{\hbar \sinh \eta }%
\right) \vbox to 24pt{}\right\} 
\]
\begin{equation}
\times P_{l}^{(\mid k+\beta _{0}\mid ,\mid k+\beta _{0}\mid )}\left( \cos
\theta _{b}\right) P_{l}^{(\mid k+\beta _{0}\mid ,\mid k+\beta _{0}\mid
)}\left( \cos \theta _{a}\right) .  \label{b31}
\end{equation}
The integral can perform by noting the equality \cite{4} 
\[
\int_{0}^{\infty }dz\frac{1}{\sinh z}e^{-\frac{M\omega }{2\hbar }%
(r_{b}+r_{a})\coth z}I_{\nu }\left( \frac{\frac{M\omega }{\hbar }\sqrt{%
r_{b}r_{a}}}{\sinh z}\right) 
\]
\begin{equation}
=\frac{1}{2}\int_{0}^{\infty }\frac{dS}{S}e^{-\frac{{\cal E}}{\hbar }%
S}e^{-m(r_{b}^{2}+r_{a}^{2})/2\hbar S}I_{\nu /2}\left( \frac{m}{\hbar }\frac{%
r_{b}r_{a}}{S}\right) ,  \label{b32}
\end{equation}
where ${\cal E}$ is defined as $(m^{2}c^{4}-E^{2})/2mc^{2}.$ We finally
obtain the exact Green's function of the relativistic three-dimensional A-B
effect 
\[
G({\bf x}_{b},{\bf x}_{a};E)=\frac{i\hbar }{2mc}\sum_{k=-\infty }^{\infty
}\sum_{l=0}^{\infty }e^{ik(\varphi _{b}-\varphi _{a})}\left( \cos \theta
_{b}/2\cos \theta _{a}/2\sin \theta _{b}/2\sin \theta _{a}/2\right) ^{\mid
k+\beta _{0}\mid } 
\]
\[
\times \frac{m}{2\pi \hbar \sqrt{r_{b}r_{a}}}\frac{l!\Gamma \left( l+2\mid
k+\beta _{0}\mid +1\right) \left( 2l+2\mid k+\beta _{0}\mid +1\right) }{%
\Gamma ^{2}\left( l+\mid k+\beta _{0}\mid +1\right) } 
\]
\[
\times \left\{ \vbox to 24pt{}I_{\nu }\left[ \sqrt{\frac{m{\cal E}}{2\hbar
^{2}}}\left( \left( r_{b}+r_{a}\right) -\left| r_{b}-r_{a}\right| \right) 
\right] K_{\nu }\left[ \sqrt{\frac{m{\cal E}}{2\hbar ^{2}}}\left( \left(
r_{b}+r_{a}\right) +\left| r_{b}-r_{a}\right| \right) \right] \vbox to 24pt{}%
\right\} 
\]
\begin{equation}
\times P_{l}^{(\mid k+\beta _{0}\mid ,\mid k+\beta _{0}\mid )}\left( \cos
\theta _{b}\right) P_{l}^{(\mid k+\beta _{0}\mid ,\mid k+\beta _{0}\mid
)}\left( \cos \theta _{a}\right)  \label{b33}
\end{equation}
with $\nu =l+\mid k+\beta _{0}\mid +1/2$. It is worth to note that there
exist no bound states in the pure A-B effect. This is reasonable, since we
are treating a scattering system.

According to the orthogonality relations of Jacobi polynomials \cite{9a}, 
\[
\int_{-1}^{-1}dx\left( 1-x\right) ^{\alpha }\left( 1+x\right) ^{\beta
}P_{n}^{(\alpha ,\beta )}\left( x\right) P_{m}^{(\alpha ,\beta )}\left(
x\right) 
\]
\begin{equation}
=\frac{2^{\alpha +\beta +1}}{\alpha +\beta +2n+1}\frac{\Gamma \left( \alpha
+n+1\right) \Gamma \left( \beta +n+1\right) }{n!\Gamma \left( \alpha +\beta
+n+1\right) }\delta _{m,n}
\end{equation}
,we find the radial Green's function of the relativistic three-dimensional
A-B effect 
\[
G(r_{b},r_{a};E)=\frac{2m}{\hbar \sqrt{r_{b}r_{a}}}
\]
\begin{equation}
\times \left\{ \vbox to 24pt{}I_{\nu }\left[ \sqrt{\frac{m{\cal E}}{2\hbar
^{2}}}\left( \left( r_{b}+r_{a}\right) -\left| r_{b}-r_{a}\right| \right) %
\right] K_{\nu }\left[ \sqrt{\frac{m{\cal E}}{2\hbar ^{2}}}\left( \left(
r_{b}+r_{a}\right) +\left| r_{b}-r_{a}\right| \right) \right] \vbox to 24pt{}%
\right\} .  \label{b34}
\end{equation}
By applying the method developed in Refs. \cite{5,5a} again, the effect of a
spherically shaped impenetrable wall located at the radius $r=a$ can be
researched via formula of Eq. (\ref{a33}). This gives ,e.g. for $%
r_{a}<r_{b}<a$, the exact Green's function$:$%
\[
G^{({\rm wall})}(r_{b},r_{a};E)=\frac{2m}{\hbar \sqrt{r_{b}r_{a}}}
\]
\begin{equation}
\times \left[ I_{\nu }\left( \sqrt{2m{\cal E}/\hbar ^{2}}a\right) K_{\nu
}\left( \sqrt{2m{\cal E}/\hbar ^{2}}r_{b}\right) -\left( K\leftrightarrow
I\right) \right] \frac{I_{\nu }\left( \sqrt{2m{\cal E}/\hbar ^{2}}%
r_{a}\right) }{I_{\nu }\left( \sqrt{2m{\cal E}/\hbar ^{2}}a\right) }.
\label{b35}
\end{equation}
The corresponding bound state energy spectra are given by the equation 
\begin{equation}
I_{l+\mid k+\beta _{0}\mid +1/2}\left( \sqrt{2m{\cal E}/\hbar ^{2}}a\right)
=0.  \label{b36}
\end{equation}
We again see that the presence of the flux line in the circular billiard
simply changes the order of the Bessel functions. The energy spectra is
determined by the zero points of the modified Bessel function. This quantum
effect may detect by the experiment. It has much interested in the mesoscope
systems \cite{1f,1g,1h,1c,1d}. For the non-relativistic quantum A-B billiard
system, the exact Green's function is given by replacing the pseudoenergy $%
{\cal E}$ with $-E.$

\section{Concluding Remarks}

In this contribution, the Green's function of the relativistic two and
three-dimensional A-B system is given by path integral approach. The results
are separated into the angular and radial parts. From the radial parts, the
Green's function of the relativistic two and three-dimensional quantum A-B
billiard system are obtained via the closed formula of Dirichlet boundary
condition given by \ the $\delta $-function perturbation. The energy spectra
are determined by the zeros of the modified Bessel function involving the
partial wave expanded Green's function of the unperturbed A-B systems. The
A-B system serve as the prototype of arbitrary systems bounded by the
spherical Dirichlet boundary condition. There is an interesting effect \cite
{1f,1g,1h} in mesoscope systems related to our results for the
non-relativistic case. Such effect can be exactly described by the A-B
magnetic field surrounded by a spherically shaped $\delta $-function. Its
Green's function is given by \cite{5,5a} 
\begin{equation}
G^{(\delta )}(r_{b},r_{a};E)=G(r_{b},r_{a};E)-\frac{G(r_{b},a;E)G(a,r_{a};E)%
}{G(a,a;E)-\hbar /\alpha a^{(D-1)}}  \label{c1}
\end{equation}
with $G$ being the radial Green's function without $\delta (r-a)$-potential
and $\alpha $ the interacting strength of $\delta $-function. In
two-dimensional case, with the result of Eq. (\ref{a36}), Eq. (\ref{c1})
yields for $r_{b}>a>r_{a}$ 
\begin{equation}
G^{(\delta )}(r_{b},r_{a};E)=-\frac{2m}{\alpha a}\frac{I_{\left| n+\beta
_{0}\right| }\left( \sqrt{-2mE/\hbar ^{2}}r_{a}\right) K_{\left| n+\beta
_{0}\right| }\left( \sqrt{-2mE/\hbar ^{2}}r_{b}\right) }{\frac{2m}{\hbar }%
I_{\left| n+\beta _{0}\right| }\left( \sqrt{-2mE/\hbar ^{2}}a\right)
K_{\left| n+\beta _{0}\right| }\left( \sqrt{-2mE/\hbar ^{2}}a\right) -\frac{%
\hbar }{\alpha a}}.  \label{c2}
\end{equation}
Energy spectra $E_{n}$ of bound states are determined by the equation 
\begin{equation}
\frac{\hbar ^{2}}{2m\alpha a}=I_{\left| n+\beta _{0}\right| }\left( \sqrt{%
-2mE_{n}/\hbar ^{2}}a\right) K_{\left| n+\beta _{0}\right| }\left( \sqrt{%
-2mE_{n}/\hbar ^{2}}a\right) .  \label{c3}
\end{equation}
From the asymptotic behavior of $I_{\alpha }\left( \alpha z\right) K_{\alpha
}\left( \alpha z\right) $ for $\alpha \rightarrow \infty $ [p. 378 \cite{9b}%
] 
\begin{equation}
I_{\alpha }\left( \alpha z\right) K_{\alpha }\left( \alpha z\right) \approx 
\frac{1}{2\alpha \sqrt{1+z^{2}}},  \label{c4}
\end{equation}
we obtain 
\begin{equation}
\frac{\hbar ^{2}}{m\alpha a}\approx \left( \left| n+\beta _{0}\right| +\frac{%
2m\left| E_{n}\right| a^{2}}{\hbar ^{2}}\right) ^{-1/2}<\frac{1}{\left|
n+\beta _{0}\right| }.  \label{c5}
\end{equation}
This implies only finite bound states exist and the upper bound is given by 
\begin{equation}
\left| n+\beta _{0}\right| <\frac{m\alpha a}{\hbar ^{2}}.  \label{c6}
\end{equation}
We see that the A-B effect not just change the energy levels but change the
number of bound states. On the other hand, the radius of a thin-walled
cylinder also affects the number of energy levels. For the three-dimensional
case, with the radial Green's function (\ref{b34}) of A-B effect, we have
the Green's function of semi-transparent wall for $r_{b}>a>r_{a}$ 
\begin{equation}
G^{(\delta )}(r_{b},r_{a};E)=-\frac{1}{\sqrt{r_{b}r_{a}}}\frac{\frac{2m}{%
\alpha a^{2}}I_{l+\left| k+\beta _{0}\right| +1/2}\left( \sqrt{-2mE/\hbar
^{2}}r_{a}\right) K_{l+\left| k+\beta _{0}\right| +1/2}\left( \sqrt{%
-2mE/\hbar ^{2}}r_{b}\right) }{\frac{2m}{\hbar a}I_{l+\left| k+\beta
_{0}\right| +1/2}\left( \sqrt{-2mE/\hbar ^{2}}a\right) K_{l+\left| k+\beta
_{0}\right| +1/2}\left( \sqrt{-2mE/\hbar ^{2}}a\right) -\frac{\hbar }{\alpha
a^{2}}}.  \label{c7}
\end{equation}
The energy levels of bound states are determined by 
\begin{equation}
\frac{\hbar ^{2}}{2m\alpha a}=I_{l+\left| k+\beta _{0}\right| +1/2}\left( 
\sqrt{-2mE_{n}/\hbar ^{2}}a\right) K_{l+\left| k+\beta _{0}\right|
+1/2}\left( \sqrt{-2mE_{n}/\hbar ^{2}}a\right) .  \label{c8}
\end{equation}
A similar analysis of two dimensional case gives the upper bound of
eigenvalue for the $\left( l+\left| k+\beta _{0}\right| +1/2\right)
\rightarrow \infty $%
\begin{equation}
\frac{\hbar ^{2}}{m\alpha a}\approx \left( l+\left| k+\beta _{0}\right| +1/2+%
\frac{2m\left| E_{n}\right| a^{2}}{\hbar ^{2}}\right) ^{-1/2}<\frac{1}{%
l+\left| k+\beta _{0}\right| +1/2},  \label{c9}
\end{equation}
i.e., 
\begin{equation}
\left( l+\left| k+\beta _{0}\right| +1/2\right) <\frac{m\alpha a}{\hbar ^{2}}%
.  \label{c10}
\end{equation}
\ It is easily to see that the number of bound states of three-dimensional
case is more than the two-dimensional one for an extra degree of freedom of
quantum number. These results give a reasonable ground in illustrating the
dependence of the electron wave functions on the vector potential of the A-B
magnetic fields while electron penetrates the spherical cylinder.

\centerline{ACKNOWLEDGMENTS} The work is supported by the National Science
Council of Taiwan under contract number NSC88-2811-M-009-0015. 

\end{document}